\documentclass[showpacs,eqsecnum,aps]{revtex4}
\usepackage{amsmath}
\usepackage{amssymb}
\usepackage{amsfonts}
\usepackage{graphicx}
\usepackage{color}
\def\beq{\begin{equation}}
\def\eeq{\end{equation}}
\def\beqarray{\begin{eqnarray}}
\def\eeqarray{\end{eqnarray}}

\def\bDelta{\mbox{\boldmath$\Delta$}}
\def\sumint{\sum\mspace{-18mu}\int}

\begin{document}
\title{A Poincar\'e invariant treatment of the three-nucleon problem.}

\author{W. Polyzou,M. Tucker, S. Veerasamy}
\affiliation{University of Iowa, Iowa City, IA 52246}

\author{Ch. Elster, T. Lin}
\affiliation{
Institute of Nuclear and Particle Physics,  and
Department of Physics and Astronomy,  Ohio University, Athens, OH 45701}

\author{W. Gl\"ockle}
\affiliation{
Institut f\"ur Theoretische Physik II, Ruhr-Universit\"at Bochum,
D-44780 Bochum, Germany}

\author{H.~Wita{\l}a,  J.~Golak, R.~Skibi\'nski}
\affiliation{M. Smoluchowski Institute of Physics, Jagiellonian University,
PL-30059 Krak\'ow, Poland}

\author{H.\ Kamada}
\affiliation{
Department of Physics, Faculty of Engineering, Kyushu Institute
of Technology,
1-1 Sensuicho Tobata, Kitakyushu 804-8550, Japan
}

\author{B. Keister}
\affiliation{
Physics Division, 1015N
National Science Foundation
4201 Wilson Blvd.
Arlington, VA  22230
}

\begin{abstract}
{I summarize recent progress in the treatment of the 
Poincar\'e three-nucleon problem at intermediate energies}
\end{abstract}

\maketitle

The essential properties of mathematical models of few-nucleon systems
that are applicable at energy scales up to a few GeV are (1) the model
is a quantum theory (2) Poincar\'e invariance is an exact symmetry of
the model (3) the energy spectrum is bounded from below and (4) the
model satisfies space-like cluster properties.  We discuss models of
the three-nucleon system with these properties and demonstrate that
these models can provide a realistic description of three-nucleon
observables at these energy scales.

The mathematical setting for a quantum theory is a Hilbert space.
Probability amplitudes are represented by Hilbert space scalar
products of unit normalized rays.  The Hilbert space in these models
is the tensor product of mass $m$ spin ${1/2}$ irreducible
representation spaces of the Poincar\'e group, where $m$ is the
nucleon mass.

The Poincar\'e group has ten infinitesimal generators.  From these
generators it is possible to construct two Casimir invariants $(m^2,
w^2=j^2/m^2)$, four additional mutually commuting observables,
$\mathbf{h}$, and 4 operators, $\bDelta \mathbf{h}$, that are conjugate
to the commuting observables $\mathbf{h}$.  The spectrum of the
commuting observables $\mathbf{h}$ is determined by the conjugate
operators $\Delta \mathbf{h}$ and the spectrum of the Casimir
operators.  For particles, the eigenvalues of the Casimir operators
$m$ and $j$ are the mass and spin of the particle.  The Hilbert space
of square integrable functions of the eigenvalues of the operators
$\mathbf{h}$ over their joint spectrum is a mass $m$ spin $j$
irreducible representation space for the Poincar\'e group.  The most
common choice for $\mathbf{h}$ is the three components of the linear
momentum and one component of the canonical spin; however, the
canonical spin could be replaced by the light-front spin, helicity, or
one component of the Pauli-Lubanski vector; the linear momentum could
be replaced by light-front components of the four momenta, four
velocity, or the Newton-Winger position operator.  We denote the
single nucleon Hilbert space by ${\cal H}_1$.

There is a natural unitary irreducible representation of the
Poincar\'e group on ${\cal H}_1$. The
Poincar\'e group Wigner ${\cal D}$-functions,
\[
{\cal D}^{jm}_{\mathbf{h}' ; \mathbf{h}}
[\Lambda ,a ] := 
\langle (m,j) \mathbf{h}' \vert U(\Lambda ,a) 
\vert (m,j) \mathbf{h} \rangle ,
\] 
in the irreducible basis $\{ \vert (m,j) \mathbf{h} \rangle \}$   
are known \cite{brad}. 

Few-nucleon Hilbert spaces are tensor products of single nucleon
spaces, ${\cal H}_1$. The tensor product of the single nucleon
irreducible representations of the Poincar\'e group defines the
kinematic representation of the Poincar\'e group on the few-nucleon
Hilbert space.  These kinematic representations are reducible; they
can be decomposed into a direct integral of irreducible
representations of the Poincar\'e group using the Poincar\'e group
Clebsch-Gordan coefficients.  The basis-dependent Poincar\'e group 
Clebsch-Gordan coefficients,
\[ 
C(1,2:3)=\langle (m_1,j_1) \mathbf{h}_1 ,
(m_2,j_2) \mathbf{h}_2 \vert (m_3,j_3) \mathbf{h}_3 ,\eta \rangle , 
\]
are known\cite{brad}.  The parameter $\eta$ represents invariant 
degeneracy quantum numbers that separate
multiple copies of irreducible representations with the same mass and
spin.  These Clebsch-Gordan coefficients satisfy
\[ 
\sumint {\cal D}^{j_1m_1}_{\mathbf{h}_1 ; \mathbf{h}_1'} [\Lambda ,a ]
{\cal D}^{j_2m_2}_{\mathbf{h}_2 ; \mathbf{h}_2'} [\Lambda ,a ]
d\mathbf{h}_1' d\mathbf{h}_2' C(1',2':3) =
\sumint C(1,2:3') d\mathbf{h}_3' {\cal
D}^{j_3m_3}_{\mathbf{h}_3' ; \mathbf{h}_3} [\Lambda ,a ] .
\]

Dynamical representations of the Poincar\'e group are constructed by
adding interactions to the non-interacting mass operator, $M_0$, that
commute with the four operators $\mathbf{h}$ that label distinct
vectors in irreducible subspaces, the four conjugate operators
$\bDelta \mathbf{h}$, and the spin.  The interacting mass operator in the
non-interacting irreducible basis has a kernel of the form
\[
\langle (m,j) \mathbf{h} ,\eta \vert M \vert (m',j') 
\mathbf{h}' ,\eta' \rangle = 
\delta (\mathbf{h} ; \mathbf{h}')\delta_{jj'} 
\langle m ,\eta \vert M^j \vert m' ,\eta'
\rangle
\]
where
$\delta (\mathbf{h} ; \mathbf{h}')$ is a product of Dirac delta functions
in the continuous variables and Kronecker delta functions in the 
discrete variables.  Diagonalizing $M$ in this basis 
\[
\sumint \langle m ,\eta \vert M^j \vert m' ,\eta'
\rangle dm'd\eta' \langle m', \eta' \vert \psi \rangle = 
\lambda  \langle m, \eta \vert \psi \rangle
\]
gives simultaneous eigenstates of
$M$, $j$, and $\mathbf{h}$.  The resulting
eigenfunctions 
\[
\langle (m,j) \mathbf{h} ,\eta \vert (\lambda ,j') \mathbf{h}' \rangle 
=
\delta (\mathbf{h};\mathbf{h}') \delta_{jj'} 
\langle m ,\eta \vert \psi \rangle 
\]
are complete, and because $\{M,j,\mathbf{h},\bDelta\mathbf{h}\}$ have
the same commutation relations as
$\{M_0,j,\mathbf{h},\bDelta\mathbf{h}\}$, the eigenstates transform
irreducibly.  The matrix elements of the irreducible representation in
this basis are identical to the Poincar\'e Wigner ${\cal D}$ functions
for the free particle representation of the same spin with the
particle mass $m$ replaced by the eigenvalue $\lambda$ of $M$.  These
Poincar\'e-Wigner ${\cal D}$ functions define all of the matrix elements of a
dynamical representation of the Poincar\'e group in this basis of
irreducible eigenstates of $M$.  The spectral condition is satisfied if the
binding energy is less than the sum of the masses of the bound
particles.

The problem of constructing realistic two-body interactions as input 
to the relativistic three-nucleon problem is solved by using 
existing realistic nucleon-nucleon interactions \cite{CDBonn}\cite{V18}
as input. If we define the relative momentum operator $\mathbf{k}^2$
as a function of the kinematic two-body mass operator by
\[
M_0(\mathbf{k}^2)  := \sqrt{\mathbf{k}^2 + m_1^2} + \sqrt{\mathbf{k}^2 + m_2^2},
\]
then a dynamical two-body mass operator is defined by 
\[
M = M_0 (\mathbf{k}^2+ 2\mu V)
\]
where $\mu$ is the two-nucleon reduced mass and $\langle (m',j')
\mathbf{h}' ,\eta' \vert V \vert (m,j) \mathbf{h} ,\eta \rangle =
\delta_{j'j}\delta (\mathbf{h}; \mathbf{h}') \langle k,\eta \vert V^j
\vert k ,\eta \rangle$, where $\langle k,\eta \vert V^j \vert k ,\eta \rangle$
is the kernel of a non-relativistic potential that is fit to
differential scattering cross section data correctly transformed to
the two-body center of momentum frame.  Time-dependent scattering
theory, along with the Kato-Birman invariance principle, can be used
to show that the M{\o}ller wave operators for the Hamiltonian
associated with the mass operator, $M$, and the non-relativistic
Hamiltonian associated with the interaction, $V$, are identical
functions of $\mathbf{k}^2$ \cite{brad}.  It follows that the
relativistic and non-relativistic $S$-matrices are identical functions
of $\mathbf{k}^2,\eta$ and ,$j$.  For nucleon-nucleon applications the
Clebsch-Gordan coefficients can be designed so the degeneracy
parameters $\eta$ have the same spectrum as the non-relativistic $l^2$ and
$s^2$ that appear as variables in typical nucleon-nucleon
interactions.  This means that high-precision interactions fit to
scattering data can be used directly in Poincar\'e invariant
two-nucleon models without modification.
 
Cluster properties require that multiparticle representations of the
Poincar\'e group $U(\Lambda ,a)$ approach tensor products of the
subsystem representations, $U_i(\Lambda ,a)$, when the subsystems are
asymptotically separated by large space-like displacements:
\begin{equation}
\lim_{(b_i-b_j)^2 \to \infty}  
\Vert (U(\Lambda ,a) - \otimes U_k (\Lambda,a) )
\prod U_l (I,b_l) \vert \psi \rangle\Vert  =0 .
\label{cluster}
\end{equation}
For a system consisting of an interacting pair of nucleons and a
spectator nucleon it is possible to first decompose the tensor product
of three irreducible representations of the Poincar\'e group into a
superposition of irreducible representations and then add the two-body
interaction.  Alternatively, it is also possible to add an interaction
directly to the two-body irreducible representation to construct an
interacting two-body representation and then decompose the tensor
product of that representation and the spectator representation into a
three-body irreducible representation.  In both cases the result is a
three-body irreducible representation of the Poincar\'e group with an
interacting pair of particles.  If the non-trivial part of the
interaction kernel, $\langle k',\eta' \vert V^j \vert k ,\eta \rangle$,
is the same in both constructions then both constructions give identical 
scattering matrices. It is also easy to demonstrate that the first 
construction fails to satisfy the cluster property (\ref{cluster}) 
while the second one satisfies property (\ref{cluster}) by construction.

The advantage of the first approach is that two-body interactions for
different pairs can be combined in a manner that preserves the
Poincar\'e symmetry.  This is because the two-body interactions for
each pair commute with the $j,\mathbf{h},\bDelta\mathbf{h}$ of the
non-interacting three-body irreducible representation.  This
construction gives a unitary representation of the Poincar\'e group for
three interacting particles that unfortunately fails to satisfy
cluster properties.  The identity of the $S$-matrices for the separate
2+1 body problems implies \cite{ekstein60} that the two $2+1$-body 
representations are related by a unitary operator, $A_{(ij)(k)}$,
called a scattering equivalence.  Cluster properties can be restored
without breaking the Poincar\'e invariance by multiplying the
three-body representation that fails to satisfy cluster properties
by the (symmetrized) \cite{coester82}\cite{brad} product of 
the unitary operators
$A_{(ij)(k)}$ for all three pairs.  Because the product of scattering
equivalences is a scattering equivalence, this transformation does
not change the on-shell three-body $S$ matrix.  It also shows that
even though the untransformed representation of the Poincar\'e group
fails to satisfy cluster properties, the resulting $S$ matrix
satisfies cluster properties.  These scattering equivalences
are only needed when the three-body solutions are used in four- or
more-body models or when they are used to compute matrix elements
of electroweak current operators.

Note that the input relativistic two-body interactions fit to
experiment and cluster properties fix the few-body dynamical operators
up to three- or more-body interactions.  Kinematic subgroups normally
associated with Dirac's forms of dynamics correspond to different
choices of $\mathbf{h}$ and $\bDelta\mathbf{h}$.  Models with a given
kinematic subgroup are scattering equivalent to models with any other
kinematic subgroup, but the scattering equivalences generate many-body
interactions under change of representation.

The operator form of the Faddeev equation for the three-body problem
is identical to the corresponding non-relativistic equation.  The
differences are (1) how the two-body interactions are embedded in the
Faddeev kernel (2) the structure of the recoupling coefficients, which
in the relativistic case are Racah coefficients of the Poincar\'e
group that change the order of the pairwise coupling of irreducible
representations (3) and kinematic factors.  Relativistic effects are
measured by comparing the difference between the relativistic and
non-relativistic three-body calculations with identical two-body
input.  Since in both cases the two-body interactions are designed to
fit experiment, there are no relativistic corrections at the two-body
level.

Technical differences in the relativistic and non-relativistic Faddeev
equation appear when the kernel and driving terms are evaluated in a
basis\cite{fritz65}\cite{brad1}.  In the relativistic case the
two-body interactions appear inside of square root operators, the
Racah coefficients for the Poincar\'e group have a non-trivial spin
and momentum dependence, and for systems of more than three particles
it is necessary to compute the scattering equivalences that restore
cluster properties.  While these properties make the relativistic
model more complicated from a computational point of view, it is
possible to treat all of these complications without approximation
\cite{walter86}\cite{cps}\cite{kamada:2007}\cite{brad1}.

For realistic interactions the Faddeev equations have been solved for
energies up to 250 MeV in a representation with a Euclidean 
kinematic subgroup using a partial wave expansion\cite{witala}
\cite{witala2}.  To extend these calculations to reactions with
energies above 250 MeV we have demonstrated that the Faddeev equations
can also be accurately solved by direct integration without partial
waves \cite{lin}\cite{lin2}\cite{lin3}.  Converged results, using
direct integration, are obtained for energies up to 2 GeV.

At low energies, calculations with realistic interactions exhibit only
small differences with the corresponding non-relativistic results
\cite{witala} .  Corrections to the Triton binding energy depend on
the two-body interaction, but generally lead to a small decrease in
the binding energy\cite{kamada}\cite{kamadaap}, typically less than a
tenth of an MeV.  The low-energy calculations also show nontrivial
contributions to the observable $A_y$ at energies (5-13 MeV)
\cite{witala2} due to relativistic spin rotations, confirming the
scale, but not the sign of an effect observed by Miller and
Schwenk\cite{miller} in a simpler model.
While these relativistic contributions
move the calculations away from the experimental results, they imply
that these effects need to be accounted for in final resolution of the
$A_y$ puzzle.

The higher-energy calculations based on direct integration used the
spin-independent Malfliet-Tjon interaction. These calculations
demonstrated that approximations that include only relativistic
kinematics can lead to large ``effects'' that are almost completely
canceled by dynamical relativistic effects.  These calculations
\cite{lin2}\cite{lin3} also demonstrated the non-uniformity of the
convergence of the multiple scattering series for both elastic
scattering and breakup reactions.  For breakup reactions away from the
quasielastic peak at least one iteration of the series was required to
obtain converged results, even above 1 GeV.  There were also
departures from the non-relativistic calculations based on
the same two-body interaction in the neighborhood of the quasielastic
peak.  Relativistic and non-relativistic exclusive breakup
calculations at 500 MeV show a different energy dependence \cite{lin2}
for different pairs of proton angles.  The behavior of the
relativistic calculations with the Malfilet-Tjon interaction seem to
explain differences between experiment and the corresponding
non-relativistic calculation \cite{lin2} for a large number of angle pairs.

To test the effects of ignoring the scattering equivalences that
restore cluster properties in four-body calculations we (1) turned off
the interaction with one of the nucleons in the three-body
construction outlined above and (2) alternatively took a tensor
product of a two-body calculation and a spectator nucleon.  The two
representations are scattering equivalent.  We assumed that the
interacting nucleons were bound in a deuteron state.  Then we
calculated the charge form factor for the electron
scattering off of the spectator proton in the presence of the free
deuteron.  In the tensor product representation the momentum of the
deuteron does not affect the form factor, however in the representation
where the two-body interaction is added to the non-interacting
three-body irreducible representation, there is a clear dependence of
the form factor on the momentum of the deuteron.  Figure 1 illustrates
the difference between these two calculations as a function of
momentum transfer and the component of the momentum of the deuteron
parallel to the momentum transfer.  This unphysical effect can be as
large as 6\% and is a consequence of not transforming to a
representation where both the current and dynamics clusters.  This
suggests that the three-nucleon wave functions should be transformed
to a representation where the dynamics satisfies cluster properties
before they can be used in the computation of electromagnetic
observables.

We also tested some modern soft interactions \cite{jisp16}\cite{n3lo}
that are precisely fit to the same scattering data as realistic
meson-exchange interactions.  These interactions are designed to be
used in low-energy many-body calculations; the transformations that
reduce the contribution of the high-momentum components to the
two-body interactions also generate many-body interactions and
exchange currents.  It is interesting to investigate the extent to
which these kinds of interactions can be used in models that study
intermediate energy dynamics without introducing the generated
three-body interactions and exchange currents.  In this case we
considered the problem of elastic electron-deuteron scattering in a
Poincar\'e covariant impulse approximation in a model with a
light-front kinematic subgroup.  We performed calculations of the
elastic scattering observables $A$, $B$ and $T_{20}$ \cite{chung}
using the AV18 \cite{V18} and CD-Bonn \cite{CDBonn} interactions with
calculations using soft interactions fit to the same two-body data.
All calculations used the same impulse current as input.  While each
calculation requires different exchange current contributions,
calculations of all three electromagnetic observables for momentum
transfers in the GeV range were well described using standard meson
exchange potentials like Argonne V18 or CD Bonn.  For these
interactions the small discrepancies with experiment in all three
observables are due to well-understood two-body exchange currents.  On
the other hand, for the calculations using the N3LO or the JISP 16
wave functions the exchange current contributions that are needed to
explain the discrepancies between calculations and experiment become
significant in the observable $A$ for momentum transfers between .5
and 1 GeV$^2$, in $B$ for momentum transfers above .5 GeV$^2$ and in
the tensor polarization for momentum transfers above .15 GeV$^2$.
Figure 2. shows the result of the four calculations of $A$ up to
$Q^2=2.5$ (GeV)$^2$.  So while these soft interactions are clearly
useful at low energy, the additional exchange currents that are
required in a relativistic calculation limit the benefits of having a
softer potential.

The research summarized above indicates that it is now possible to
extend the few-nucleon physics program that has been successful at
low energies to treat few-nucleon problems at intermediate
energy scales in a manner that respects all fundamental principles
of physics that are relevant at these scales.

\clearpage

\begin{figure}
\begin{center}
\rotatebox{0}{
\includegraphics[width=15cm]{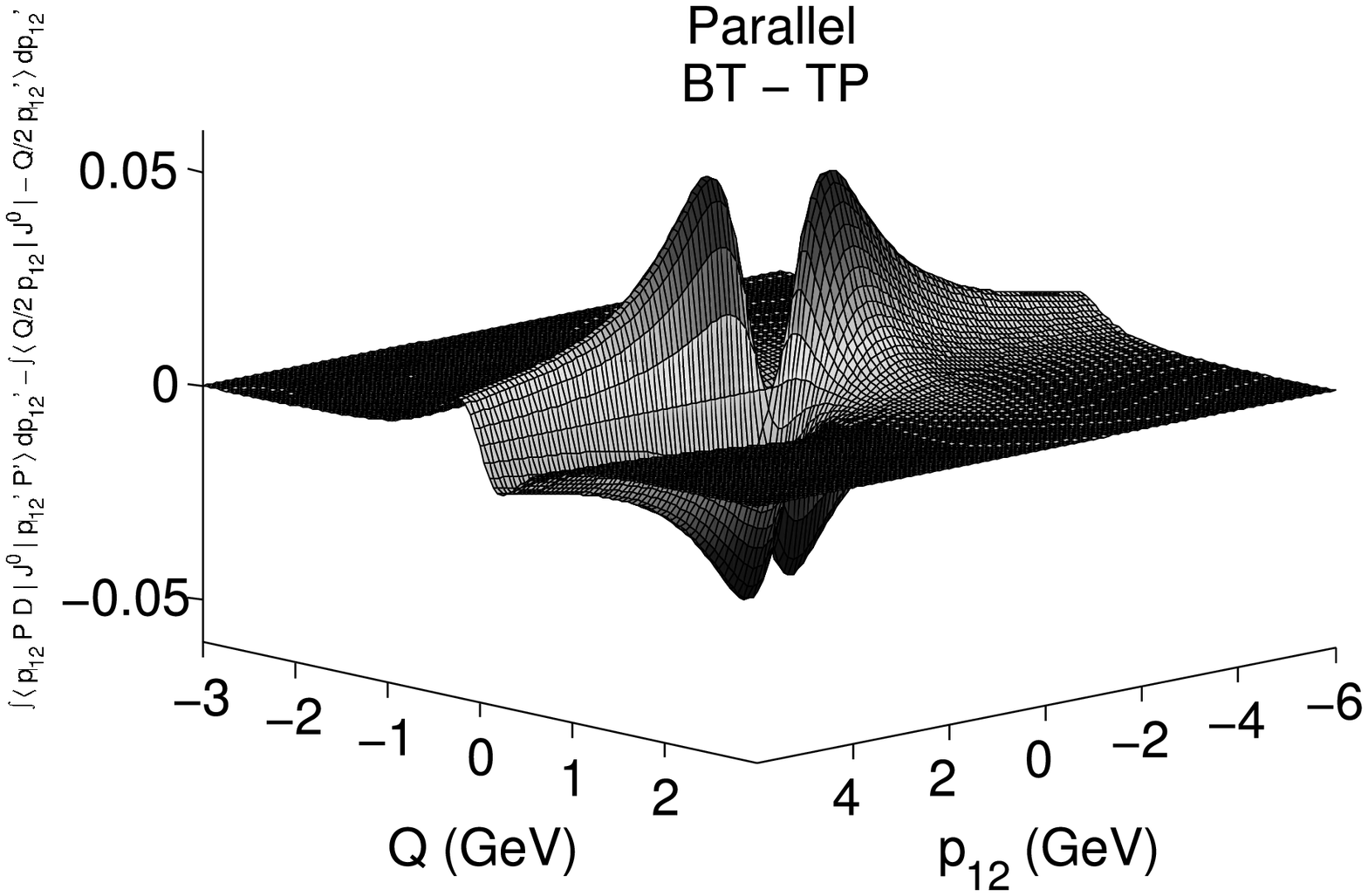}
}
\end{center}
\caption{Comparison of charge 
form factors with and without corrections for cluster properties as a 
function of spectator momentum
\label{fig1}
}
\end{figure}

\clearpage

\begin{figure}
\begin{center}
\rotatebox{270}{
\includegraphics[width=15cm]{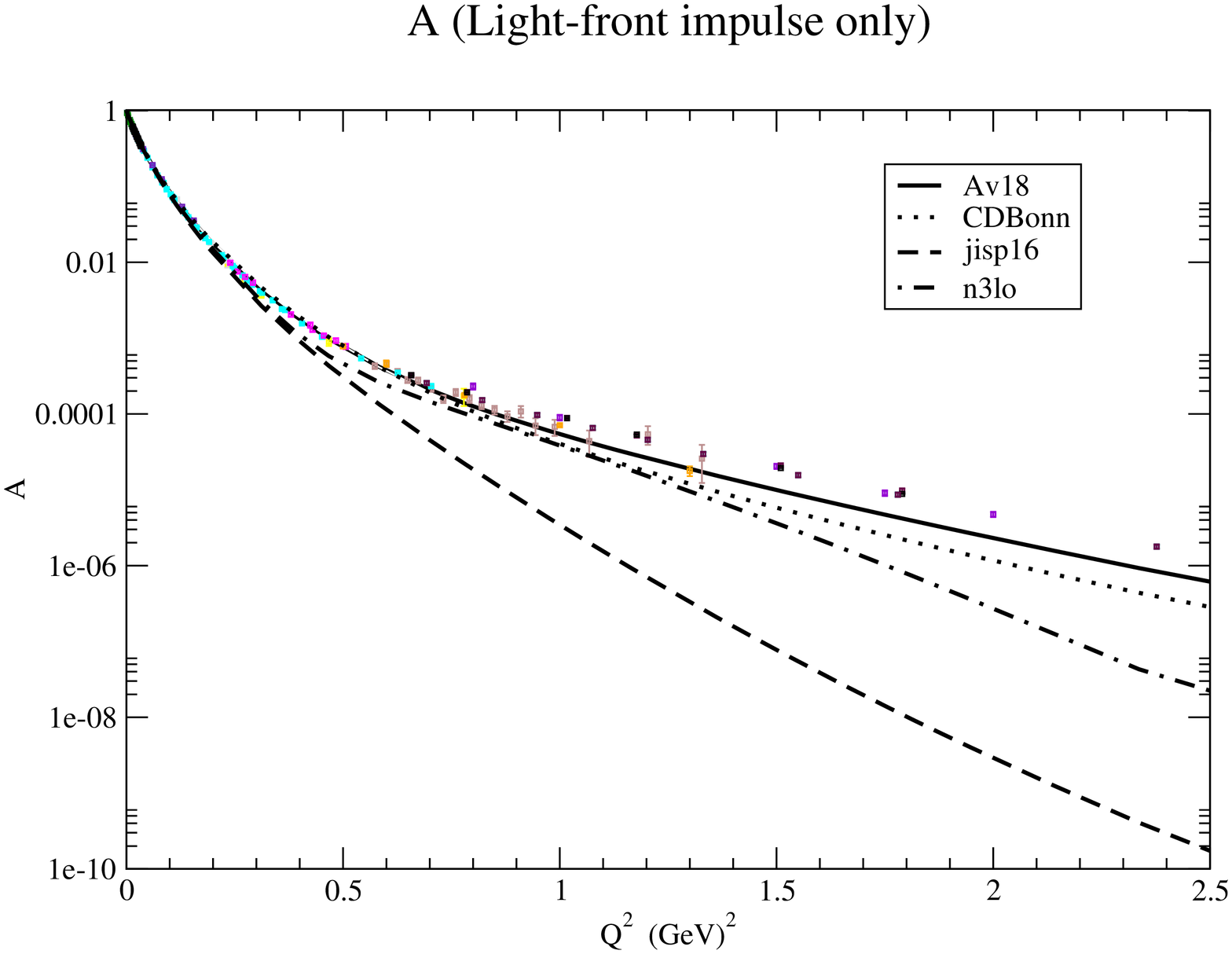}
}
\end{center}
\caption{Deuteron 
structure function $A$ in the impulse approximation for different 
hard and soft nucleon-nucleon interactions.
\label{fig2}
}
\end{figure}

\end{document}